\documentclass[preprint,10pt,authoryear,5p, twocolumn, times]{elsarticle}

\usepackage{amssymb}
\usepackage{amsmath}
\usepackage{hyperref}
\usepackage{bbm}
\usepackage[table,xcdraw,dvipsnames]{xcolor}

\usepackage{tikz}
\usetikzlibrary{positioning}
\definecolor{color1}{rgb}{0.0, 0.4470, 0.7410}
\definecolor{color2}{rgb}{0.8500, 0.3250, 0.0980}
\definecolor{color3}{rgb}{0.9290, 0.6940, 0.1250}
\definecolor{color4}{rgb}{0.4940, 0.1840, 0.5560}

\DeclareMathOperator*{\argmin}{argmin}

\journal{Energy Economics}

\begin{document}

\begin{frontmatter}

\title{Postprocessing of point predictions for probabilistic forecasting of day-ahead electricity prices: The benefits of using Isotonic Distributional Regression}

\author[inst1]{Arkadiusz Lipiecki}

\affiliation[inst1]{organization={Doctoral School, Faculty of Management},
    addressline={Wroc{\l}aw University of Science and Technology}, 
            city={Wroc{\l}aw},
            postcode={50-370},
            country={Poland}}
\author[inst2]{Bartosz Uniejewski}
\author[inst2]{Rafa{\l} Weron}

\affiliation[inst2]{organization={Department of Operations Research and Business Intelligence},
        addressline={Wroc{\l}aw University of Science and Technology}, 
            city={Wroc{\l}aw},
            postcode={50-370},
            country={Poland}}

\begin{abstract}
Operational decisions relying on predictive distributions of electricity prices can result in significantly higher profits compared to those based solely on point forecasts. However, the majority of models developed in both academic and industrial settings provide only point predictions. To address this, we examine three postprocessing methods for converting point forecasts of day-ahead electricity prices into probabilistic ones: Quantile Regression Averaging, Conformal Prediction, and the recently introduced Isotonic Distributional Regression. We find that while the latter demonstrates the most varied behavior, it contributes the most to the ensemble of the three predictive distributions, as measured by Shapley values. Remarkably, the performance of the combination is superior to that of state-of-the-art Distributional Deep Neural Networks over two 4.5-year test periods from the German and Spanish power markets, spanning the COVID pandemic and the war in Ukraine. 
\end{abstract}


\end{frontmatter}

\section{Introduction}
Recent studies demonstrate that operational decisions based on probabilistic price forecasts can lead to significantly (up to 20\%) higher profits in day-ahead electricity trading than those relying solely on point predictions \citep{uni:wer:21,mar:nar:wer:zie:23}. However, constructing models that can yield probabilistic forecasts is a complex task. 
No wonder the majority of methods developed by both academics and practitioners provide only point predictions \citep{now:wer:18,zie:ste:18}. A workable solution is to use so-called \textit{postprocessing} to convert point forecasts into probabilistic ones \citep{che:jan:ste:ler:24,van:etal:21:postprocessing}, as such an approach can benefit from developments in the point forecasting literature \citep{liu:now:hon:wer:17}.

In this study, we compare an established postprocessing method in energy forecasting -- \textit{Quantile Regression Averaging} \cite[QRA;][]{liu:now:hon:wer:17,wan:etal:19,kat:zie:21,uni:wer:21,nit:wer:23,yan:yan:liu:23} -- with \textit{Conformal Prediction} \cite[CP;][]{sha:vov:08,kat:zie:21}, popular in the machine learning community, and the recently introduced \textit{Isotonic Distributional Regression} \cite[IDR;][]{hen:zie:gne:21,gne:ler:sch:23}. 
Since we are not interested in developing point forecasting models for day-ahead markets, but rather in employing point predictions as inputs to postprocessing schemes, we use a variant of the well-performing \textit{LASSO-Estimated AutoRegressive} (LEAR) model of \cite{lag:mar:sch:wer:21}, and a simple similar-day `naive' benchmark, commonly used as a reference point in \textit{electricity price forecasting} \cite[EPF;][]{wer:14}. 
The obtained predictive distributions are compared to three probabilistic benchmarks built on point forecasts of the LEAR or the naive model and normally N($0,\hat\sigma$) distributed errors, as well as state-of-the-art Distributional Deep Neural Networks \cite[DDNNs;][]{mar:nar:wer:zie:23}. 
Two major European electricity markets -- Germany and Spain -- serve as our testing ground. 

The remainder of the paper is structured as follows. In Section \ref{sec:Datasets} we present the datasets, then in Section \ref{sec:Point:forecasts} we explain how the point forecasts of day-ahead electricity prices are computed. Next, in Section \ref{sec:Postprocessing} we describe the three postprocessing schemes: Quantile Regression Averaging, Conformal Prediction, and Isotonic Distributional Regression. In Section \ref{sec:Results} we first briefly recall the Continuous Ranked Probability Score (CRPS), then discuss the obtained results in terms of the CRPS and the test for Conditional Predictive Ability (CPA) of \cite{gia:whi:06}. Next, we use Shapley values~\citep{cov:lun:lee:20:SAGE,lun:20:LossSHAP} to see which component contributes the most to the ensemble of the three predictive distributions. We conclude Section \ref{sec:Results} by taking a risk management perspective and presenting results for the tails of the predictive distribution. Finally, in Section \ref{sec:Conclusions} we summarize the main results.

\begin{figure*}
 \centering
 \includegraphics[width=.85\linewidth]{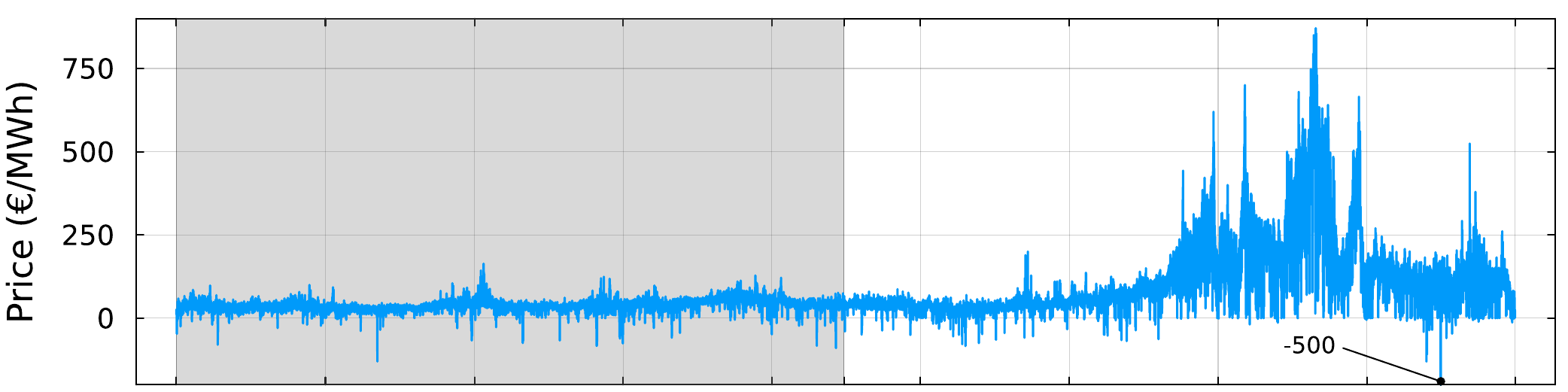}
 \includegraphics[width=.85\linewidth]{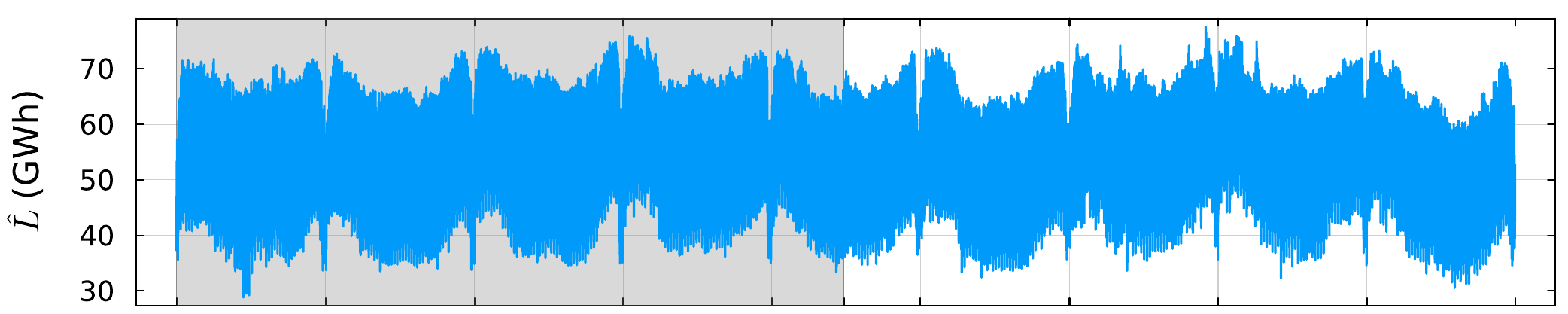}
 \includegraphics[width=.85\linewidth]{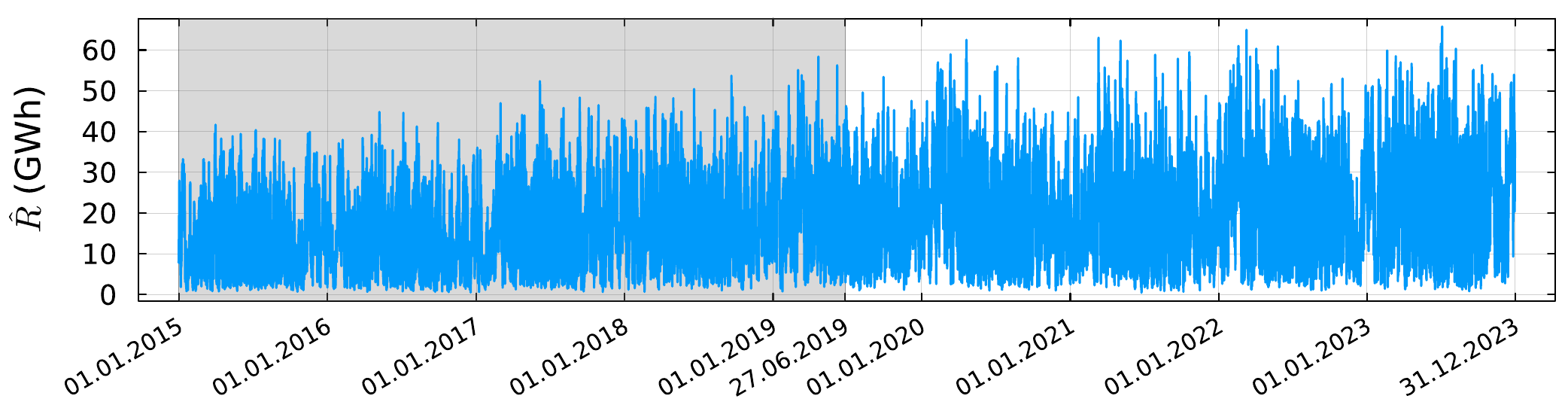}
 \includegraphics[width=.85\linewidth]{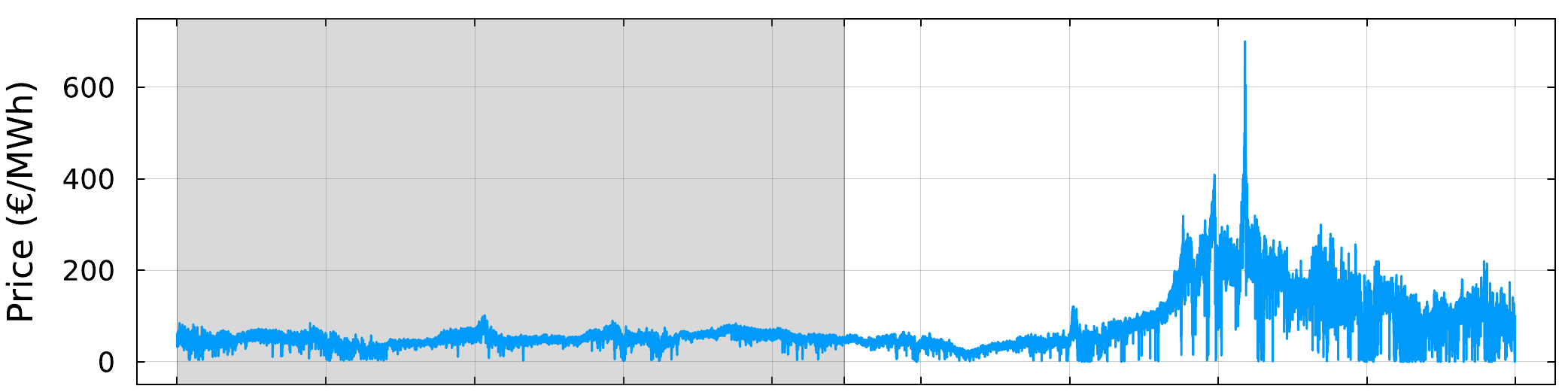}
 \includegraphics[width=.85\linewidth]{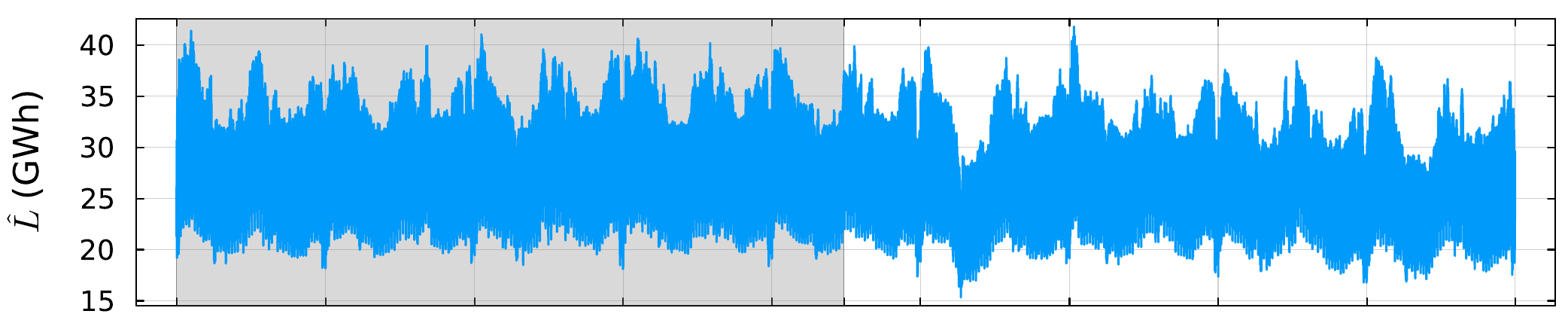}
 \includegraphics[width=.85\linewidth]{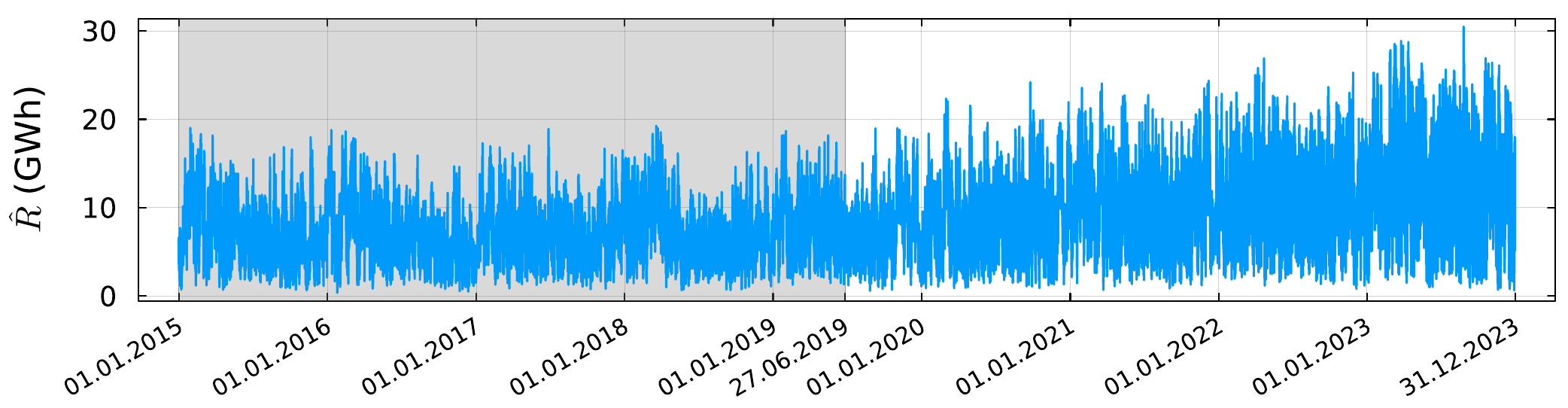}
 \caption{\textit{From top to bottom}: Day-ahead electricity prices $p_{d,h}$ and day-ahead predictions of load $\hat{L}_{d,h}$ and renewable generation $\hat{R}_{d,h}$ (onshore/offshore wind and solar) in Germany (\textit{top three panels}) and Spain (\textit{bottom three panels}). Gray background marks the initial calibration window (1.01.2015-26.06.2019), while white corresponds to the 4.5-year test period (27.06.2019-31.12.2023). 
 Note that in Germany the prices can be negative and on 02.07.2023 the day-ahead price dropped to the minimum admissible level of $-500$ EUR/MWh.}
\label{fig:data}
 \end{figure*}

\section{Datasets}
\label{sec:Datasets}

The data we use is publicly available and has been downloaded from ENTSO-E (\url{https://transparency.entsoe.eu}; day-ahead prices, day-ahead load forecasts, day-ahead onshore/offshore wind and solar generation forecasts) and Investing.com (\url{https://www.investing.com/}; carbon emission, natural gas, crude oil and coal closing prices). More precisely, the German day-ahead prices are for the BZN$|$DE-LU bidding zone (BZN$|$DE-AT-LU until 30.09.2018) and the Spanish day-ahead prices for the BZN$|$ES bidding zone. The day-ahead load forecasts and \textit{renewable energy sources} (RES) generation forecasts are for the two countries -- DE and ES, respectively. Since some of the ENTSO-E data has a 15-minute resolution, we have aggregated it to hourly values. The European Union Allowance (EUA) carbon emission prices, natural gas prices from the Title Transfer Facility (TTF) virtual trading point in the Netherlands, Brent crude oil prices, and API2 coal prices are the last known closing prices at the time of bidding in the day-ahead market.

Datasets for both markets span from 01.01.2015 to 31.12.2023; the 4.5-year out-of-sample test periods start on 27.06.2019, see Fig.~\ref{fig:data}. 
All time series were preprocessed to account for transitions to/from daylight saving time (DST). Missing values, which occur during the switch to DST, were replaced with the arithmetic average of the observations from the surrounding hours. Duplicate values, which occur during the switch back, were replaced by their arithmetic mean.

\section{Computing point forecasts}
\label{sec:Point:forecasts}

\subsection{The LEAR model}
\label{ssec:LEAR}

We use a variant of the \textit{LASSO-Estimated AutoRegressive} (LEAR) model of \cite{lag:mar:sch:wer:21} to generate high quality point forecasts $\hat{p}_{d,h}$ of day-ahead electricity prices for day $d$ and hour $h$. It is a parameter-rich autoregressive structure with exogenous variables estimated using the \textit{Least Absolute Shrinkage and Selection Operator} \cite[LASSO;][]{has:tib:wai:15}. 
In the original formulation, the regressors include past prices $\boldsymbol{p}_{d-k} = \{p_{d-k,1}, ..., p_{d-k,24}\}$ for lags $k=1,2,3,7$, day-ahead predictions 
$\boldsymbol{x}^{(i)}_{d-k} = \{{x}^{(i)}_{d-k,1}, ..., {x}^{(i)}_{d-k,24}\}$ of two ($i=1,2$) fundamental variables for lags $k=0,1,7$, and daily dummies to capture the weekly seasonality. The LEAR model we use has the form: 
\begin{align}\label{eq:LEAR}
    p_{d,h} =& \sum_{h=1}^{24} \beta_h p_{d-1,h} + \sum_{h=1}^{24} 
    \beta_{h+24} p_{d-2,h} + \nonumber \\ 
    &+ \sum_{h=1}^{24} \beta_{h+48} p_{d-3,h} + \sum_{h=1}^{24} \beta_{h+72} p_{d-7,h} \nonumber \\ 
    &+ \sum_{h=1}^{24} \beta_{h+96} \hat{L}_{d,h} + \sum_{h=1}^{24} \beta_{h+120} \hat{L}_{d-1,h} + \sum_{h=1}^{24} \beta_{h+144} \hat{L}_{d-7,h} \nonumber \\
    &+ \sum_{h=1}^{24} \beta_{h+168} \hat{R}_{d,h} 
    + \sum_{h=1}^{24} \beta_{h+192} \hat{R}_{d-1,h} 
    + \sum_{h=1}^{24} \beta_{h+216} \hat{R}_{d-7,h}
    \nonumber \\ 
    &+ \beta_{241} \text{EUA}_{d-2} + \beta_{242} \text{NG}_{d-2} + \beta_{243} \text{Brent}_{d-2} \nonumber \\ 
    & + \beta_{244} \text{API2}_{d-2} + \sum_{i=1}^7 \beta_{i+244} D_i + \varepsilon_{d,h},
\end{align}
since, following \cite{mar:nar:wer:zie:23}, we:
\begin{itemize}
    \setlength\itemsep{0em}
    \item use day-ahead predictions of the system-wide load $\hat{L}_{d-k,h}$ as ${x}^{(1)}_{d-k,1}$ and day-ahead RES (sum of onshore/offshore wind and solar) generation $\hat{R}_{d-k,h}$ as ${x}^{(2)}_{d-k,1}$;
    \item additionally include four macroeconomic variables that have a major impact on European electricity prices: EUA$_{d-2}$ carbon emission prices, NG$_{d-2}$  natural gas prices, 
    Brent$_{d-2}$ crude oil prices, and API2$_{d-2}$ coal prices; all four are the last known closing prices on day $d-2$.
\end{itemize}

Moreover, like \cite{lag:mar:sch:wer:21} and \cite{zie:wer:18} but unlike \cite{mar:nar:wer:zie:23}, we 
preprocess the electricity prices with the \textit{area hyperbolic sine} variance stabilizing transformation: 
\begin{equation}\label{eq:asinh}
   \mbox{asinh}(x) = \log\left(x +\sqrt{x^2+1} \right),
\end{equation}
where $x$ is the price standardized by subtracting the in-sample median and dividing by the median absolute deviation (MAD), adjusted by the 75\% quantile of the standard normal distribution for asymptotical consistency with the standard deviation \citep{uni:wer:zie:18}. To recover price forecasts we apply the inverse transformation, i.e., the hyperbolic sine, to the generated predictions; see \cite{nar:zie:19} for a more accurate back-transformation. 

Finally, unlike \cite{lag:mar:sch:wer:21} and \cite{mar:nar:wer:zie:23}, instead of using the faster but less accurate Least Angle Regression \cite[LARS;][]{efr:has:joh:tib:04}, we use the standard coordinate descent LASSO estimator \cite[as implemented in Matlab 2024a; see][]{fri:has:tib:10} to estimate the model coefficients. We combine the latter with 7-fold cross-validation (CV), like \cite{mar:nar:wer:zie:23} but unlike \cite{lag:mar:sch:wer:21} who used the Akaike Information Criterion (AIC) for initial estimation and coordinate descent for the final run. Since in our setup CV 
involves a random split of the training data, resulting in a slightly different forecast for each run, we compute the predictions of the LEAR model not once, but 5 times for each training window length and average the 5 individual results to obtain the final LEAR forecast. 
As we will see in Table \ref{tab:res}, these changes, compared to the variant used by \cite{mar:nar:wer:zie:23}, result in more accurate point forecasts, leading to significantly better predictive distributions.

We consider a rolling window setup, where forecasts of all 24 hours on day $d$ are calculated in the morning of day $d-1$ and the model parameters are reestimated each day using a calibration sample of $D$ most recent past observations.
As in the original LEAR formulation, the parameters are estimated separately for each of the 4 training window lengths $D = 56, 84, 1092$ and $1456$, yielding point forecasts $\hat{p}_{d,h}^{56}$, $\hat{p}_{d,h}^{84}$, $\hat{p}_{d,h}^{1092}$ and $\hat{p}_{d,h}^{1456}$ for each day and hour in the test period.

\subsection{The naive benchmark}

As a reference point, we use a popular in EPF implementation of the similar-day approach, often called the \textit{naive} method \citep{lag:mar:sch:wer:21,wer:14,zie:wer:18}. It uses last week's prices to forecast the prices on Monday, Saturday and Sunday, and yesterday's prices for the remaining days:
\begin{equation}\label{eq:naive}
\hat{p}_{d,h} = \begin{cases}
    p_{d-7,h},  & \text{for $d=$ Mon, Sat or Sun}, \\
    p_{d-1,h}, & \text{otherwise}.
\end{cases}
\end{equation}

\section{Postprocessing point forecasts}
\label{sec:Postprocessing}

Our goal is to obtain 99 percentiles of the predictive distribution $\hat{F}_{p}$ of $p_{d,h}$. 
Like for the LEAR model, we construct separate distributional models for each hour $h$ and retrain them daily, using 4 different calibration windows with the most recent data: $\{\left(\hat{p}_{t,h}, p_{t, h}\right)\}_{t = d-m}^{d-1}$ with $m \in \{28,56,91,182\}$. Note that we use the term `training/calibration window' to refer to the data used to estimate the point/probabilistic forecasting model. For postprocessing we use the open source Julia package \nobreak{\textit{PostForecasts.jl}} \citep{postforecasts:github}.

Before fitting distributional models, we must first generate point forecasts for $182$ days, and thus our approach requires a total of 1456 (longest training window) + 182 (longest calibration window) = 1638 days to generate the first predictive distribution. The final probabilistic forecasts are obtained via probability \cite[or `vertical';][]{lic:g-c:win:13} averaging of 4 distributions obtained for different calibration window lengths $m$; except for the Naive-1N model which uses only the 182-day window, see Section \ref{ssec:normal:bench} for details. Let us now briefly describe the postprocessing schemes.

\subsection{Quantile Regression Averaging (QRA)} 

Formally introduced by \cite{now:wer:15}, and successfully used in the GEFCom2014 competition \citep{mac:now:16,gai:gou:ned:16} and later energy forecasting applications \citep{liu:now:hon:wer:17,wan:etal:19,kat:zie:21,uni:wer:21,nit:wer:23,yan:yan:liu:23,cor:din:pou:24}, the method estimates conditional quantiles of the target variable as a linear combination of point predictions in a quantile regression setting:
\begin{equation}
    \hat{q}(\alpha|\boldsymbol{\hat{p}}_{d, h}) = [1,\boldsymbol{\hat{p}}_{d, h}]~\mathbf{\beta}_\alpha,
\end{equation}
where $\hat{q}(\alpha|\cdot)$ is the conditional $\alpha$th quantile, $\boldsymbol{\hat{p}}_{d, h}$ is the row vector of point predictions (see the next paragraph for details), and $\boldsymbol{\beta}_\alpha$ is the column vector of coefficients.
Prediction intervals (PIs) are obtained by running QRA for two selected quantiles, e.g., the 5\% and 95\% quantiles yield the 90\% PI. 
The coefficients are computed by minimizing the pinball score, so to obtain $\hat{F}_{p}$, a linear optimization problem must be solved independently for each quantile \citep{now:wer:18}.  
This makes QRA by far the most computationally intensive method we consider, yet still feasible on a consumer-grade laptop. 

Given the pool of four point forecasts $\hat{p}_{d,h}^{56}, \hat{p}_{d,h}^{84}, \hat{p}_{d,h}^{1092}, \hat{p}_{d,h}^{1456}$, we initially examined three approaches: 
\begin{enumerate}
    \item[(i)] using a single model with all individual point forecasts as regressors $\boldsymbol{\hat{p}}_{d, h} = [\hat{p}_{d,h}^{56}, \hat{p}_{d,h}^{84}, \hat{p}_{d,h}^{1092}, \hat{p}_{d,h}^{1456}]$, like in the original formulation of QRA \citep{now:wer:15}; 
    \item[(ii)] using a single model with one regressor being the average point forecast $\boldsymbol{\hat{p}}_{d, h}= \hat{p}_{d,h}^{ave}=\frac14(\hat{p}_{d,h}^{56}+ ... +\hat{p}_{d,h}^{1456})$, a variant dubbed Quantile Regression Machine (QRM) in \cite{mar:uni:wer:20IJF}; and 
    \item[(iii)] averaging (over quantiles or probabilities) the predictive distributions $\hat{F}_{p}^{56}, ..., \hat{F}_{p}^{1456}$ obtained from the individual point forecasts $\hat{p}_{d,h}^{56}, ..., \hat{p}_{d,h}^{1456}$, respectively. 
\end{enumerate}
Below we present the results for approach (ii), which turned out to be the fastest and the most accurate. We call it the \textbf{LEAR-QRM} model.

\subsection{Conformal Prediction (CP)} 

This is a framework for computing PIs based on absolute point prediction errors in a chosen calibration window \citep{sha:vov:08}. CP produces valid intervals for a given confidence level and requires no distributional assumptions \citep{zaf:etal:22}. However, the estimated PIs are centered on the point forecast, so obtaining quantile forecasts from CP is only possible under the assumption of symmetrically distributed errors. The $\alpha$th quantile is given by:
\begin{equation}\label{eq:CP}
    \hat{q}(\alpha|\hat{p}_{d, h}) = \begin{cases}
      \hat{p}_{d, h} - \lambda^{2\alpha} & \text{if}\quad \alpha < 1/2, \\
      \hat{p}_{d, h} + \lambda^{2(1-\alpha)} & \text{otherwise},
    \end{cases}     
\end{equation}
where $\lambda^{\alpha}$ is the so-called \textit{nonconformity score} such that $[\hat{p}_{d, h} - \lambda^\alpha,  \hat{p}_{d, h} + \lambda^\alpha]$ is a ($1-\alpha$) PI.
We use the same inductive scheme as \cite{kat:zie:21}, but take a different approach to selecting training and calibration sets. Although the rolling windows we use are not disjoint, each point prediction error calculated in a calibration window does not belong to the training window used for generating that particular prediction. 
Similar to QRA, we present the results for CP computed for $\hat{p}_{d,h} = \hat{p}_{d,h}^{ave} =\frac14(\hat{p}_{d,h}^{56}+ ... +\hat{p}_{d,h}^{1456})$ in Eq.\ \eqref{eq:CP}; we call it the \textbf{LEAR-CP} model. This approach outperformed combining 
the predictive distributions $\hat{F}_{p}^{56}, ..., \hat{F}_{p}^{1456}$.

\subsection{Isotonic Distributional Regression (IDR)}

This is a nonparametric method for learning conditional distributions under the stochastic order constraint \citep{hen:zie:gne:21,wal:hen:zie:gne:24}. The output $\hat{F}_p$ minimizes the Continuous Ranked Probability Score (CRPS; see Sec.\ \ref{ssec:Results:CRPS})  under the isotonic constraint, which requires that the conditional cumulative distribution function (CDF) of the response be non-increasing (or equivalently, that the quantiles of the response be non-decreasing) with respect to the regressor. This makes postprocessing point forecasts a natural setting for IDR, since a forecast of the response variable generally satisfies the isotonic relation. 
For a calibration window of $m$ data points $(p_{i,h}, \hat{p}_{i,h})_{i=d-m,...,d-1}$, renumbered $(p_{i,h}, \hat{p}_{i,h})_{i=1,...,m}$ so that $\hat{p}_{1,h} \leq ... \leq \hat{p}_{m,h}$, IDR estimates $m$ conditional distributions $\hat{F}_{i}(z) \equiv \hat{F}(z|\hat{p}_{i, h})$:
\begin{equation}
    \left(\hat{F}_1(z),...,\hat{F}_m(z)\right) = \argmin_{(\eta_1,...,\eta_m)} \sum_{i=1}^m \left(\eta_i - \mathbbm{1}_{\{p_{i, h} \leq z\}}\right)^2,
    \label{eqn:idrminmax}
\end{equation}
with $\eta_1 \geq ... \geq \eta_m$ and $\eta_i \in [0,1]$. 
To obtain the conditional distribution for any $\hat{p}_{d,h}\in\mathbbm{R}$, we adopt the interpolation method suggested in \cite{hen:zie:gne:21}:
\begin{equation}
    \hat{F}_d(z) = \frac{\hat{p}_{d, h}-\hat{p}_{i, h}}{\hat{p}_{i+1, h} - \hat{p}_{i, h}}\hat{F}_i(z) + \frac{\hat{p}_{i+1, h} - \hat{p}_{d, h}}{\hat{p}_{i+1, h} - \hat{p}_{i, h}} \hat{F}_{i+1}(z),
    \label{eqn:idrinterp}
\end{equation}
for any $i \in \{1,\,...,\,m-1\}$ such that $\hat{p}_{d, h}\in[\hat{p}_{i, h}, \hat{p}_{i+1}]$; if $\hat{p}_{d, h} < \hat{p}_{1, h}$ or $\hat{p}_{d, h} > \hat{p}_{m, h}$, we set  
$\hat{F}_d(z) = \hat{F}_1(z)$ or $\hat{F}_d(z) = \hat{F}_m(z)$, 
respectively.

\begin{figure}[t!]
\centering 
\begin{tikzpicture}[scale=1]

\node at (-0.6,5.1) {\large \textbf{(a)}};

\draw[-stealth, black, very thick] (0, 0) -- (5, 0);
\draw[-stealth, black, very thick] (0, 0) -- (0, 5);

\draw[gray] (1.5, 0) -- (1.5, 1) node [pos=0, below left = 1pt and -12pt, black] {$\hat{p}_{1,h}~$};
\draw[gray] (2, 0) -- (2, 2.5) node [pos=0, below left = 1pt and -12pt, black] {$~\hat{p}_{2,h}$};
\draw[gray] (3.5, 0) -- (3.5, 2) node [pos=0, below left = 1pt and -12pt, black] {$\hat{p}_{3,h}$};
\draw[gray] (4.5, 0) -- (4.5, 4) node [pos=0, below left = 1pt and -12pt, black] {$\hat{p}_{4,h}$};

\draw[gray] (0, 1) -- (1.5, 1) node [pos=0, above left = -8pt and 1pt, black] {$p_{1,h}$};
\draw[gray] (0, 2.5) -- (2, 2.5) node [pos=0, above left = -8pt and 1pt, black] {$p_{2,h}$};
\draw[gray] (0, 2) -- (3.5, 2) node [pos=0,above left = -8pt and 1pt, black] {$p_{3,h}$};
\draw[gray] (0, 4) -- (4.5, 4) node [pos=0,above left = -8pt and 1pt, black] {$p_{4,h}$};

\filldraw[black] (1.5, 1) circle (2pt);
\filldraw[black] (2, 2.5) circle (2pt);
\filldraw[black] (3.5, 2) circle (2pt);
\filldraw[black] (4.5, 4) circle (2pt);

\filldraw[rotate around={-20:(2.75, 2.25)}, color=color2, opacity = 0.3] (2.75, 2.25) ellipse (1.2 and 0.5);

\node at (-0.6,-0.9) {\large \textbf{(b)}};

\draw[-stealth, black, very thick] (0, -4) -- (5, -4) node [pos = 1, below right, black] {$z$};
\draw[-stealth, black, very thick] (0, -4) -- (0, -1) node [pos=0, above left = -7pt and 1pt, black] {$0$};

\draw[gray] (1, -4) -- (1, -1.5) node [pos=0, below left = 1pt and -12pt, black] {$p_{1,h}$};
\draw[gray] (2.5, -4) -- (2.5, -1.5) node [pos=0, below left = 1pt and -12pt, black] {$~p_{2,h}$};
\draw[gray] (2, -4) -- (2, -1.5) node [pos=0, below left = 1pt and -12pt, black] {$p_{3,h}~$};
\draw[gray] (4, -4) -- (4, -1.5) node [pos=0, below left = 1pt and -12pt, black] {$p_{4,h}$};

\draw[gray] (0, -1.5) -- (4, -1.5) node [pos=0, above left = -7pt and 1pt, black] {$1$};

\draw[gray] (0, -1.5-1.25) -- (4, -1.5-1.25);
\node at (0, -1.5-1.35) [above left = -8pt and 1pt, black] {\large $\frac{1}{2}$};

\filldraw[color1] (1, -1.5) circle (4pt);
\filldraw[color1] (2, -1.5) circle (4pt);
\filldraw[color1] (2.5, -1.5) circle (4pt);
\filldraw[color1] (4, -1.5) circle (4pt) node [below right=-8pt and 6pt, color1] {$\hat{F}_1(z) \equiv \hat{F}(z|\hat{p}_{1,h})$};

\filldraw[color2] (1, -4) circle (3pt);
\filldraw[color2] (2, -1.5-1.25) circle (3pt);
\filldraw[color2] (2.5, -1.5) circle (3pt);
\filldraw[color2] (4, -1.5) circle (3pt) node [below right=6pt and 6pt, color2] {$\hat{F}_2(z), \hat{F}_3(z)$}; 

\filldraw[color3] (1, -4) circle (2pt);
\filldraw[color3] (2, -4) circle (2pt);
\filldraw[color3] (2.5, -4) circle (2pt);
\filldraw[color3] (4, -1.5) circle (2pt) node [below right=8+6+6pt and 6pt, color3] {$\hat{F}_4(z)$};

\node at (-0.6,-4.9) {\large \textbf{(c)}};

\draw[-stealth, black, very thick] (0, -8) -- (5, -8) node [pos = 1, below right, black] {$\hat{p}$};
\draw[-stealth, black, very thick] (0, -8) -- (0, -5) node [pos=0, above left = -7pt and 1pt, black] {$0$};

\draw[gray] (1.5, -8) -- (1.5, -5.5) node [pos=0, below left = 1pt and -12pt, black] {$\hat{p}_{1,h}~$};
\draw[gray] (2, -8) -- (2, -5.5) node [pos=0, below left = 1pt and -12pt, black] {$~\hat{p}_{2,h}$};
\draw[gray] (3.5, -8) -- (3.5, -5.5) node [pos=0, below left = 1pt and -12pt, black] {$\hat{p}_{3,h}$};
\draw[gray] (4.5, -8) -- (4.5, -5.5) node [pos=0, below left = 1pt and -12pt, black] {$\hat{p}_{4,h}$};

\node at (0, -5.5) [above left = -7pt and 1pt, black] {$1$};

\draw[gray] (0, -5.5-1.25) -- (4.8, -5.5-1.25);
\node at (0, -5.5-1.35) [above left = -8pt and 1pt, black] {\large $\frac{1}{2}$};

\draw[color1, line width = 1pt] (0, -5.53) -- (1.5, -5.53);
\draw[color1, line width = 1pt] (1.5, -5.53) -- (2, -8.01);
\draw[color1, line width = 1pt] (2, -8.01) -- (3.5, -8.01);
\draw[color1, line width = 1pt] (3.5, -8.01) -- (4.5, -8.01);
\draw[color1, line width = 1pt] (4.5, -8.01) -- (4.8, -8.01);

\draw[color2, line width = 1pt] (0, -5.51) -- (1.5, -5.51);
\draw[color2, line width = 1pt] (1.5, -5.51) -- (2, -5.5-1.25);
\draw[color2, line width = 1pt] (2, -5.5-1.25) -- (3.5, -5.5-1.25);
\draw[color2, line width = 1pt] (3.5, -5.5-1.25) -- (4.5, -8);
\draw[color2, line width = 1pt] (4.5, -8) -- (4.8, -8);

\draw[color4, line width = 1pt] (0, -5.49) -- (1.5, -5.49);
\draw[color4, line width = 1pt] (1.5, -5.49) -- (2, -5.49);
\draw[color4, line width = 1pt] (2, -5.49) -- (3.5, -5.49);
\draw[color4, line width = 1pt] (3.5, -5.49) -- (4.5, -7.98);
\draw[color4, line width = 1pt] (4.5, -7.98) -- (4.8, -7.98);

\draw[color3, line width = 1pt] (0, -5.47) -- (1.5, -5.47);
\draw[color3, line width = 1pt] (1.5, -5.47) -- (2, -5.47);
\draw[color3, line width = 1pt] (2, -5.47) -- (3.5, -5.47);
\draw[color3, line width = 1pt] (3.5, -5.47) -- (4.5, -5.47);
\draw[color3, line width = 1pt] (4.5, -5.47) -- (4.8, -5.47);

\draw[gray, dashed] (1, -5.3) -- (1, -8) node [pos=0, above left = 0pt and -12pt, gray] {$\hat{p}_{d',h}~$};
\draw[gray, dashed] (2.5, -5.3) -- (2.5, -8) node [pos=0, above left = 0pt and -12pt, gray] {$~\hat{p}_{d'',h}~$};
\draw[gray, dashed] (3.75, -5.3) -- (3.75, -8) node [pos=0, above left = 0pt and -12pt, gray] {$~~\hat{p}_{d''',h}~$};

\node at (5.5, -5.5) [color1, below right=0pt and -10pt] {$\hat{F}(p_{1,h}|\hat{p})$};
\node at (5.5, -5.5) [color2, below right=14pt and -10pt] {$\hat{F}(p_{3,h}|\hat{p})$};
\node at (5.5, -5.5) [color4, below right=28pt and -10pt] {$\hat{F}(p_{2,h}|\hat{p})$};
\node at (5.5, -5.5) [color3, below right=42pt and -10pt] {$\hat{F}(p_{4,h}|\hat{p})$};

\node at (-0.6,-8.9) {\large \textbf{(d)}};

\draw[-stealth, black, very thick] (0, -12) -- (5, -12) node [pos = 1, below right, black] {$p$};
\draw[-stealth, black, very thick] (0, -12) -- (0, -9) node [pos=0, above left = -7pt and 1pt, black] {$0$};

\node at (0, -9.5) [above left = -7pt and 1pt, black] {$1$};

\draw[gray] (0, -9.5-1.25) -- (4.8, -9.5-1.25);
\node at (0, -9.5-1.35) [above left = -8pt and 1pt, black] {\large $\frac{1}{2}$};

\draw[gray] (1, -12) -- (1, -9.5) node [pos=0, below left = 1pt and -12pt, black] {$p_{1,h}$};
\draw[gray] (2.5, -12) -- (2.5, -9.5) node [pos=0, below left = 1pt and -12pt, black] {$~p_{2,h}$};
\draw[gray] (2, -12) -- (2, -9.5) node [pos=0, below left = 1pt and -12pt, black] {$p_{3,h}~$};
\draw[gray] (4, -12) -- (4, -9.5) node [pos=0, below left = 1pt and -12pt, black] {$p_{4,h}$};

\draw[color1, line width = 1pt] (0, -9.48) -- (4.8, -9.48);

\draw[color2, line width = 1pt] (0, -12) -- (2, -12);
\draw[color2, line width = 1pt] (1.99, -12) -- (1.99, -9.5-1.25);
\draw[color2, line width = 1pt] (2, -9.5-1.25) -- (2.5, -9.5-1.25);
\draw[color2, line width = 1pt] (2.49, -9.5-1.25) -- (2.49, -9.5);
\draw[color2, line width = 1pt] (2.5, -9.5) -- (4.8, -9.5);

\draw[color3, line width = 1pt] (0, -12.02) -- (2, -12.02);
\draw[color3, line width = 1pt] (2.01, -12.02) -- (2.01, -9.5-1.25 - 1.25*0.25);
\draw[color3, line width = 1pt] (2, -9.5-1.25 - 1.25*0.25) -- (2.52, -9.5-1.25 - 1.25*0.25);
\draw[color3, line width = 1pt] (2.52, -9.5-1.25 - 1.25*0.25) -- (2.52, -9.5 - 1.25*0.5);
\draw[color3, line width = 1pt] (2.52, -9.5 - 1.25*0.5) -- (4., -9.5 - 1.25*0.5);
\draw[color3, line width = 1pt] (4., -9.5 - 1.25*0.5) -- (4., -9.53);
\draw[color3, line width = 1pt] (4., -9.53) -- (4.8, -9.53);

\node at (5.5, -9.5) [color1, below right=0pt and -10pt] {$\hat{F}(p|\hat{p}_{d', h})$};
\node at (5.5, -9.5) [color2, below right=14pt and -10pt] {$\hat{F}(p|\hat{p}_{d'', h})$};
\node at (5.5, -9.5) [color3, below right=28pt and -10pt] {$\hat{F}(p|\hat{p}_{d''', h})$};

\end{tikzpicture}

\caption{Schematic representation of the IDR algorithm. Panel (a) shows the calibration set of $m=4$ days for hour $h$, with price predictions $\hat{p}_{1,h} \leq \hat{p}_{2,h} \leq \hat{p}_{3,h} \leq \hat{p}_{4,h}$ and respective prices $p_{1,h}, ..., p_{4,h}$. 
Panel (b) displays the conditional CDFs after pooling together the two observations which violate the isotonic constraint (orange ellipse). Panel (c) shows $\hat{F}(z|\hat{p})$ as a function of $\hat{p}$, interpolated using Eq.\ \eqref{eqn:idrinterp}. 
The predictive distribution for the next day is obtained from the intersections of $\hat{F}(p_{i,h}|\hat{p})$ and a vertical line at the next day's point forecast. In panel (d) we depict $\hat{F}_p$ corresponding to three hypothetical next day's forecasts: $\hat{p}_{d',h}, \hat{p}_{d'',h}$ and $\hat{p}_{d''',h}$.
}
\label{fig:IDR}
\end{figure}

To solve Eq.\ \eqref{eqn:idrminmax}, we use the abridged pool-adjacent violators algorithm \citep{hen:2022}. This is illustrated in Fig.\ \ref{fig:IDR} for a sample calibration set $(p_{i, h}, \hat{p}_{i, h})_{i=1,...,4}$ of $m=4$ days for hour $h$. The four price forecasts are sorted to satisfy: $\hat{p}_{1,h} \leq ... \leq \hat{p}_{4,h}$. But then the respective prices $p_{1,h}, ..., p_{4,h}$ are not, since $p_{2,h} > p_{3,h}$. This requires pooling together the two observations which violate the isotonic constraint, see the orange ellipse in panel (a), with $\frac12$ probability mass assigned to $p_{2,h}$ and $\frac12$ to $p_{3,h}$. 
The resulting conditional CDFs are plotted in panel (b); note that $\hat{F}_{i}(z) \equiv \hat{F}(z|\hat{p}_{i, h})$ are defined only for $z\in \{p_{1,h},...,p_{4,h}\}$. Clearly, $\hat{F}_2(z)$ and $\hat{F}_3(z)$ overlap due to the pooling. On the other hand, as shown in panel (c), $\hat{F}(p_{2,h}|\hat{p})$ and $\hat{F}(p_{3,h}|\hat{p})$ as a function of $\hat{p}$ do not overlap. In this panel, the values for $\hat{p}_{d,h}\notin\{\hat{p}_{1,h},...,\hat{p}_{4,h}\}$ are interpolated using Eq.\ \eqref{eqn:idrinterp}. 
Finally, the predictive distribution $\hat{F}_p$ for the next day's forecast $\hat{p}_{5,h}$ is obtained from the intersections of $\hat{F}(p_{i,h}|\hat{p})$ and a vertical line at $\hat{p} = \hat{p}_{5,h}$. In panel (c) we use dashed lines to indicate three hypothetical next day's forecasts: $\hat{p}_{5',h}$, $\hat{p}_{5'',h}$ and $\hat{p}_{5''',h}$. The corresponding predictive distributions are plotted in panel (d).

We separately solve Eq.\ \eqref{eqn:idrminmax} for each of the four point prediction models to obtain $\hat{F}_p^{56}, \hat{F}_p^{84}, \hat{F}_p^{1092}$ and $\hat{F}_p^{1456}$. Then, we take a simple `vertical' average to obtain the \textbf{LEAR-IDR} model: 
\begin{equation}
    \hat{F}_p(z) =  \tfrac14 \left(\hat{F}_p^{56}(z) +  \hat{F}_p^{84}(z) + \hat{F}_p^{1092}(z) + \hat{F}_p^{1456}(z) \right).
\end{equation}
We can do this since the distributions are trained on the same set of prices $p_{1,h}, ..., p_{m,h}$, see panel (d) in Fig.\ \ref{fig:IDR}. 
We also tested a variant of IDR that used an average point forecast as a regressor, but it performed significantly worse.

\subsection{Ensemble of predictive distributions}
\label{ssec:LEAR-Ave}

Since combining forecasts typically improves the results \citep{oli:cha:mar:wer:dub:23,nit:wer:23}, not only in electricity price forecasting \citep{bar:ler:18,g-c:jos:20}, we consider an ensemble of the LEAR-QRM, LEAR-CP and LEAR-IDR predictive distributions and call it \textbf{LEAR-Ave}. We compute it as an average over probabilities \cite[`vertical';][]{mar:uni:wer:20IJF} of the three distributions. We also tested combinations of any two predictive distributions, but their performance was worse than of the LEAR-Ave.

\subsection{N(0,$\hat\sigma$)-based benchmarks}
\label{ssec:normal:bench}

A common assumption underlying time series models is that the innovations are Gaussian. Under this assumption, probabilistic forecasts can be obtained by computing the standard deviation $\hat\sigma$ of the prediction errors $\epsilon_{d,h}=p_{d,h}-\hat{p}_{d,h}$ in the calibration sample, then taking appropriate quantiles of the N(0,$\hat\sigma$) distribution and adding them to the point forecast $\hat{p}_{d,h}$ for the target day and hour \citep{now:wer:18}. We use this approach to construct three benchmark models:
\begin{itemize}
    \setlength\itemsep{0em}
    \item \textbf{Naive-1N} which uses point forecasts of the naive model defined in Eq.\ \eqref{eq:naive} and estimates $\hat\sigma$ on one calibration window of $m=182$ days;
    \item \textbf{Naive-N} which uses point forecasts of the naive model defined in Eq.\ \eqref{eq:naive} and estimates $\hat\sigma$ on calibration windows of $m \in \{28,56,91,182\}$ days;
    \item \textbf{LEAR-N} which uses point forecasts of the LEAR model defined in Sec.\ \ref{ssec:LEAR} and estimates $\hat\sigma$ on calibration windows of $m \in \{28,56,91,182\}$ days.
\end{itemize}

\begin{table}[tb]
\caption{Computational time required for each model/component to generate forecasts for a 4.5 year test period on a server equipped with an AMD EPYC 7713 64-core processor and 256 GB of RAM. For comparison, hyperparameter optimization for DDNN-JSU would take 2-3 weeks, as estimated for 16 times smaller hyperparameter sets (128 instead of 2048 elements).}
\label{tab:times}
\footnotesize
\begin{center}
\begin{tabular}{lr}
\hline
Model/component  & Time\\
\hline
LEAR (4 training windows, 5 runs, Matlab 2024a) & 2:30-3:00 hours \\
QRA (4 calibration windows, Julia 1.10) & 10-15 min\\
CP (4 calibration windows, Julia 1.10) & 10-15 sec\\
IDR (4 calibration windows, Julia 1.10) & 15-20 sec\\
LEAR-Ave (all components, Julia 1.10) & 2:45-3:15 hours\\
DDNN-JSU (4 networks, TensorFlow 2.9) & 6:00-6:30 hours \\
\hline
\end{tabular}
\end{center}
\end{table}

\subsection{The DDNN-JSU benchmark}
\label{ssec:DDNN:bench}

The fourth benchmark is the DDNN-JSU-pEns model of \cite{mar:nar:wer:zie:23}; we refer to it as \textbf{DDNN-JSU}. It is based on a Distributional Deep Neural Network architecture that outputs four parameters of Johnson's SU distribution. To estimate the DDNN-JSU model and obtain day-ahead price forecasts, we use the Python codes available on GitHub: \url{https://github.com/gmarcjasz/distributionalnn}.

For the whole 4.5-year German and Spanish out-of-sample test sets, we use the hyperparameter set optimized by \cite{mar:nar:wer:zie:23} for German data over the period 1.01.2015-31.12.2018; the files are available on GitHub. The rationale for this approach is provided by \cite{mar:20}, who showed that hyperparameters optimized for one electricity market can be effectively used in another one. Optimizing the hyperparameter set more frequently and for both markets could lead to better predictions, but is extremely time consuming.  A single hyperparameter optimization run takes weeks even on multi-core computing servers, see Tab.\ \ref{tab:times} and the DDNN documentation on GitHub.

\section{Empirical results}
\label{sec:Results}

\subsection{Comparison in terms of the CRPS}
\label{ssec:Results:CRPS}

The Continuous Ranked Probability Score \cite[CRPS;][]{gne:raf:07} is a proper scoring rule and the standard metric for evaluating probabilistic forecasts \citep{bil:gia:del:rav:23,mar:nar:wer:zie:23,now:wer:18}. It is defined as: 
\begin{equation}
    \operatorname{CRPS}(\hat{F}, x) = \int_{-\infty}^{\infty} \left(\hat{F}(y) - \mathbbm{1}_{\lbrace x \leq y \rbrace} \right)^2 dy,
\end{equation}
where $\hat{F}$ is the predictive distribution and $x$ is the observation, e.g., electricity price $p_{d,h}$. It can be approximated by:\footnote{Note that the scaling factor of 2 in Eqn. \eqref{eqn:CRPS:approx} is usually omitted in practice \citep{nit:wer:23}. This is also the case here.}
\begin{equation}\label{eqn:CRPS:approx}
    \operatorname{CRPS}(\hat{F}, x) \approx \frac{2}{M} \sum_{i=1}^M \operatorname{PS}\left(\hat{q}, x, q_i\right),
\end{equation}
where $\left(q_1, \ldots, q_M\right)$ is an equidistant monotonically increasing dense grid of probabilities, e.g., the 99 percentiles, $\hat{q} \equiv \hat{F}^{-1}(q)$ is the quantile forecast for quantile level $q\in(0,1)$, and
\begin{equation}
    \operatorname{PS}(\hat{q}, x, q) = \left(\mathbbm{1}_{\lbrace x < \hat{q} \rbrace} - q \right)\left(\hat{q} - x\right)
\end{equation}
is the so-called \textit{pinball score}, also known as the \textit{pinball loss}, \textit{quantile loss} or \textit{check
function} \citep{ber:zie:23,g-c:etal:17,mac:uni:wer:23}.  

\begin{table}[tb]
\caption{Continuous Ranked Probability Scores (CRPS; i.e., Aggregate Pinball Score across all 99 percentiles, compare with Tab.\ \ref{tab:res:extreme}) for the considered models and markets. Cells are colored independently for each row and market.
The test period labeled `2000$^\dag$' spans from 27.06.2019 to 31.12.2020 (554 days), the remaining three span full years (365 days). Note, that \cite{mar:nar:wer:zie:23} reported a CRPS of 1.662 for the LEAR-QRM model and 1.304 for the DDNN-JSU model in the first 554-day test period for Germany; see text for details and discussion.}
\label{tab:res}
\footnotesize
\begin{center}
\begin{tabular}{lrrrr}
\hline
Model    & 2020$^\dag$ & 2021 & 2022 & 2023\\
\hline
         & \multicolumn{4}{c}{\textit{Germany}}\\
Naive-1N & \cellcolor[HTML]{E67C73}3.548 & \cellcolor[HTML]{E67C73}9.494 & \cellcolor[HTML]{E67C73}25.346 & \cellcolor[HTML]{E67C73}12.078 \\
Naive-N  & \cellcolor[HTML]{E77F72}3.488 & \cellcolor[HTML]{E78072}9.322 & \cellcolor[HTML]{E77E72}25.064 & \cellcolor[HTML]{E98471}11.464 \\
LEAR-N   & \cellcolor[HTML]{FFD666}1.408 & \cellcolor[HTML]{F5D469}4.370 & \cellcolor[HTML]{F9D568}10.878 & \cellcolor[HTML]{FFD666}4.641  \\
LEAR-QRM & \cellcolor[HTML]{ACC878}1.350 & \cellcolor[HTML]{ADC878}4.189 & \cellcolor[HTML]{C3CC73}10.651 & \cellcolor[HTML]{AEC978}4.422  \\
LEAR-CP  & \cellcolor[HTML]{D5CF6F}1.369 & \cellcolor[HTML]{FFD666}4.399 & \cellcolor[HTML]{F5D468}10.864 & \cellcolor[HTML]{F2D369}4.582  \\
LEAR-IDR & \cellcolor[HTML]{FFD566}1.422 & \cellcolor[HTML]{FDD567}4.389 & \cellcolor[HTML]{FFD666}10.926 & \cellcolor[HTML]{8AC380}4.336  \\
LEAR-Ave & \cellcolor[HTML]{57BB8A}1.310 & \cellcolor[HTML]{57BB8A}3.970 & \cellcolor[HTML]{57BB8A}10.199 & \cellcolor[HTML]{57BB8A}4.215  \\
DDNN-JSU & \cellcolor[HTML]{9BC67C}1.342 & \cellcolor[HTML]{FBC568}5.395 & \cellcolor[HTML]{FBC768}13.375 & \cellcolor[HTML]{FDCF67}5.265  \\
\hline
         & \multicolumn{4}{c}{\textit{Spain}}\\
Naive-1N & \cellcolor[HTML]{E67C73}2.110 & \cellcolor[HTML]{E67C73}7.373 & \cellcolor[HTML]{E67C73}12.553 & \cellcolor[HTML]{E67C73}8.999  \\
Naive-N  & \cellcolor[HTML]{E88072}2.065 & \cellcolor[HTML]{E88172}7.201 & \cellcolor[HTML]{E88172}12.287 & \cellcolor[HTML]{E88172}8.806  \\
LEAR-N   & \cellcolor[HTML]{FFD566}1.018 & \cellcolor[HTML]{F2D369}4.166 & \cellcolor[HTML]{FFD666}7.412  & \cellcolor[HTML]{FFD666}4.735  \\
LEAR-QRM & \cellcolor[HTML]{BBCB75}0.976 & \cellcolor[HTML]{B4CA76}4.034 & \cellcolor[HTML]{95C57D}7.136  & \cellcolor[HTML]{FFD666}4.723  \\
LEAR-CP  & \cellcolor[HTML]{FFD566}1.014 & \cellcolor[HTML]{FFD666}4.208 & \cellcolor[HTML]{F6D468}7.371  & \cellcolor[HTML]{F9D568}4.699  \\
LEAR-IDR & \cellcolor[HTML]{D5CF6F}0.986 & \cellcolor[HTML]{F8D468}4.179 & \cellcolor[HTML]{CCCD71}7.268  & \cellcolor[HTML]{57BB8A}4.361  \\
LEAR-Ave & \cellcolor[HTML]{57BB8A}0.938 & \cellcolor[HTML]{57BB8A}3.832 & \cellcolor[HTML]{57BB8A}6.983  & \cellcolor[HTML]{5ABB8A}4.369  \\
DDNN-JSU & \cellcolor[HTML]{DDD06E}0.989 & \cellcolor[HTML]{FCCA67}4.627 & \cellcolor[HTML]{FBC768}8.299  & \cellcolor[HTML]{5FBC89}4.379 \\
\hline
\end{tabular}
\end{center}
\end{table}

\begin{figure*}[t]
\centering
\includegraphics[width=.9\linewidth]{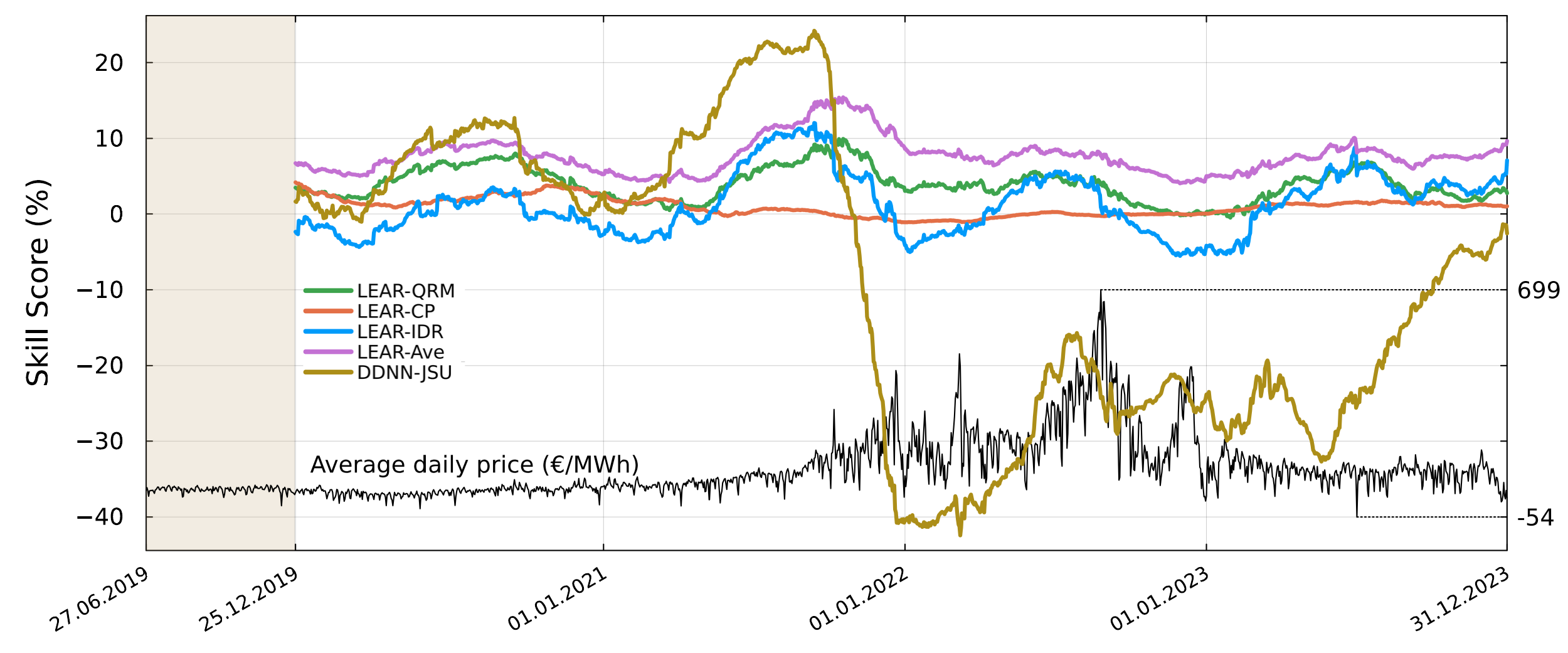}
\includegraphics[width=.9\linewidth]{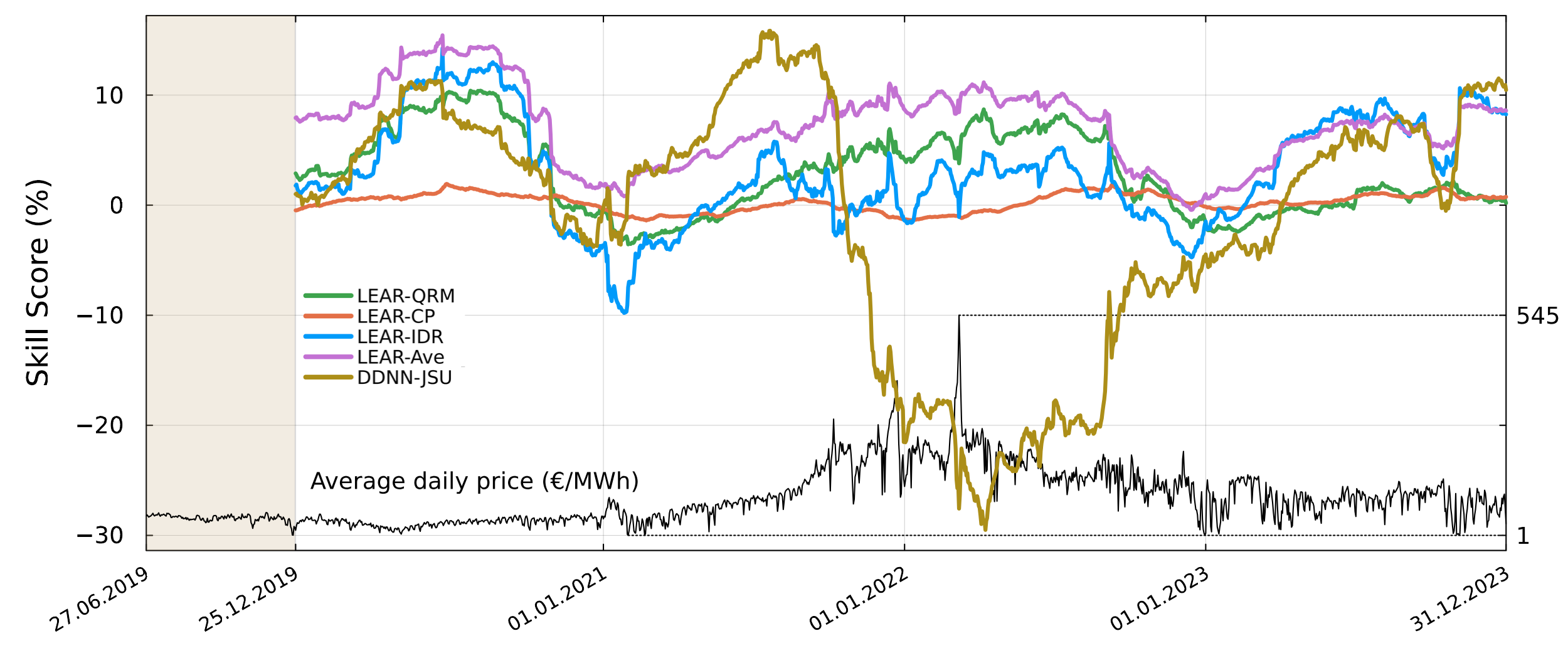}
\caption{Rolling 182-day Skill Scores, see Eq.\ \eqref{eq:skill}, for Germany (\textit{top}) and Spain (\textit{bottom}) with respect to the LEAR-N model. Values for 25.12.2019 are the scores for 27.06-25.12.2019, i.e., the beige shaded area, values for 26.12.2019 are the scores for 28.06-26.12.2019, etc. The black curves in both panels are the mean daily electricity prices $p_d = \frac{1}{24} \sum_{h} p_{d,h}$ in the depicted period; the maximum and minimum values (in EUR/MWh) are shown on the right axis.}
\label{fig:SS:CRPS}
\end{figure*}

In Table \ref{tab:res} we report the CRPS for the considered models and markets. As can be seen, the LEAR-Ave ensemble yields the lowest CRPS across both markets and all four test subperiods, while the Naive-1N and Naive-N benchmarks are the worst. Of the latter two, Naive-N significantly outperforms Naive-1N at the 5\% level for all subperiods and both markets, as measured by the CPA test of \cite{gia:whi:06}, see Sec.\ \ref{ssec:Results:GW}. This underscores the importance of estimating $\hat\sigma$ on calibration windows of different lengths, see Sec.\ \ref{ssec:normal:bench}.

The LEAR-N benchmark is much more competitive than Naive-based benchmarks due to much more accurate point forecasts. Interestingly, in some subperiods it even outperforms some of the other LEAR-based competitors. On the other hand, the DDNN-JSU model is a disappointment. It is much worse than all LEAR-based models during the energy crisis and the initial phase of the war in Ukraine (2021-2022), and performs well only in the first subperiod labeled `2000$^\dag$' in Germany and in last year in Spain. In the latter case, the LEAR-IDR model is the best performer and the second best in 2023 in Germany. It seems that LEAR-IDR excels in (relatively) calm periods that follow more volatile ones, but is not an all-rounder like LEAR-QRM.

\subsubsection{Temporal performance}

In Figure \ref{fig:SS:CRPS} we plot the rolling 182-day CRPS-based Skill Score \cite[SS; see][]{ras:ler:2018} with respect to the LEAR-N model, i.e., the LEAR model with normally N($0,\hat\sigma$) distributed errors:
\begin{equation}
\text{SS}^{model}_{d}= 1 - \frac{\sum_{k=0}^{181} \sum_{h=1}^{24} \text{CRPS}^{model}_{d-k,h}}{\sum_{k=0}^{181} \sum_{h=1}^{24} \text{CRPS}^{\text{LEAR-N}}_{d-k,h}},
\label{eq:skill}
\end{equation}
where $d=25.12.2019,...,31.12.2023$ and $\text{CRPS}^{model}_{d,h}$ is the CRPS of $model$ for day $d$ and hour $h$.

Among the three postprocessing schemes, IDR  shows the most uneven performance. In Germany relatively poor for the 182-day windows ending between 
Dec 2019 and Apr 2021, between Dec 2021 and Apr 2022, and between Aug 2022 and Apr 2023, while relatively good for the remaining periods. In Spain relatively poor for the 182-day windows ending between Nov 2020 and Apr 2021, and between Oct and Dec 2022, while relatively good for the remaining periods, especially after May 2023.
For windows that span periods of moderately increasing prices after calm periods (e.g., May-Sep 2021 in Germany and Spain) or normal prices after a spiky period (e.g., Sep-Dec 2023 in Germany and May-Dec 2023 in Spain), the IDR average significantly outperforms the QRA and CP averages. 
The CP scheme gives a relatively stable performance, both for the individual calibration windows and the average, while the QRA approach shows an intermediate behavior. 
Analyzing the four distributions $\hat{F}_p^{56}, \hat{F}_p^{84}, \hat{F}_p^{1092}$ and $\hat{F}_p^{1456}$ corresponding to the individual calibration windows of $m=28, 56, 91$ and $182$ days, the IDR-generated ones are the most volatile and different from each other, while the CP-generated ones are the least; not depicted in Fig.\ \ref{fig:SS:CRPS}.

\begin{figure}
\centering
\includegraphics[width=\linewidth]{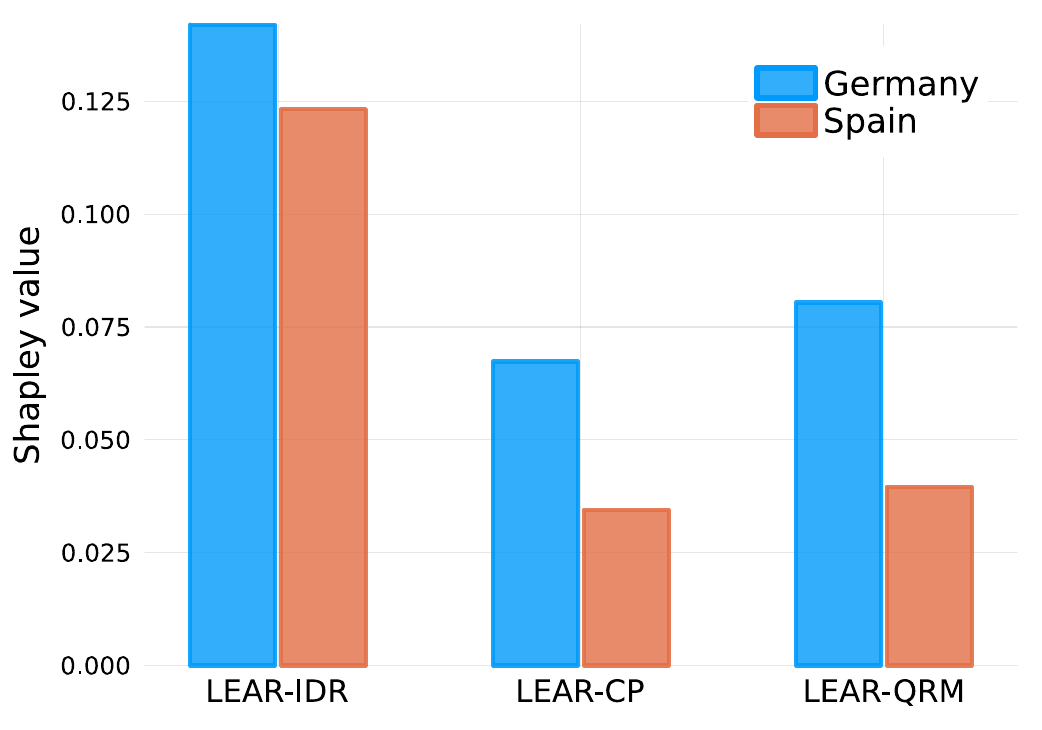}
\caption{Shapley values for the whole 4.5-year test sets in Germany and Spain. Across both markets, IDR contributes the most to the LEAR-Ave ensemble.}
\label{fig:SHAP}
\end{figure}

\subsubsection{Shapley values and component contribution to the ensemble}

We use Shapley values to assess which component contributes the most to the ensemble of the three predictive distributions in the LEAR-Ave model. Recall, that Shapley values were originally developed to fairly distribute total wins ($\rightarrow$ predictive power) among players ($\rightarrow$ ensemble components) in a cooperative game based on their individual contributions. Our approach is similar in spirit to \textit{Loss SHapley Additive exPlanations} \cite[LossSHAP;][]{lun:20:LossSHAP} and \textit{Shapley Additive Global importancE} \cite[SAGE;][]{cov:lun:lee:20:SAGE}, which aim to explain the contribution of features to the model's performance measured by a given loss function.

In Figure \ref{fig:SHAP} we plot Shapley values based on the CRPS loss -- see \citep{postforecasts:github} for details -- for the whole 4.5-year test sets in Germany and Spain. Clearly, across both markets, IDR contributes the most to the LEAR-Ave ensemble, while CP the least. 
When analyzed independently for the four subperiods (`2000$^\dag$', 2021, 2022 and 2023), the contribution of IDR is by far the highest in all but the first subperiod in Germany, when all three postprocessing schemes contribute approximately equally. In 2023 the contribution of IDR exceeds 75\% in both markets.

\subsubsection{Comparison with the results of Marcjasz et al. (2023)}

The test period labeled `2000$^\dag$' spans from 27.06.2019 to 31.12.2020 (554 days) and is the same as considered for Germany by \cite{mar:nar:wer:zie:23}. Interestingly, the latter article reported a CRPS of 1.662 for the LEAR-QRM model and 1.304 for the DDNN-JSU model for Germany. The much better performance of the LEAR-QRM model in our study (CRPS = 1.350) is a result of the improvements discussed in Sec.\ \ref{ssec:LEAR}: the use of the asinh variance stabilizing transformation \citep{uni:wer:zie:18} and the more time consuming, but more accurate coordinate descent LASSO estimator \citep{fri:has:tib:10} combined with 7-fold cross-validation.

The differences in the results of the DDNN-JSU model -- a CRPS of 1.304 vs.\ 1.342 in our study -- are harder to explain. We use exactly the same set of hyperparameters and the same Python code to estimate the weights of the neural network and make the predictions. The difference in the CRPS of $1.342-1.304=0.038$ cannot only be attributed to the stochastic nature of the estimation process; a limited simulation study suggests that this randomness could be responsible for a $\pm0.01$ discrepancy, but not more. The answer is surprising and lies in the dataset. \cite{mar:nar:wer:zie:23} used load and RES generation forecasts that are not consistent with those currently available on ENTSO-E for the years 2015-2017, and which we use in this study. In particular, the series of RES forecasts exhibit differences of considerable magnitude in both directions -- for individual hours, a median deviation of 372 MWh or ca.\ $\pm3\%$ with respect to the median RES level in the years 2015-2017, and a maximum deviation of 6,388 MWh or ca.\ $\pm45\%$!

\begin{figure}[tb]
	\centering 
	\includegraphics[width = 0.4\textwidth]{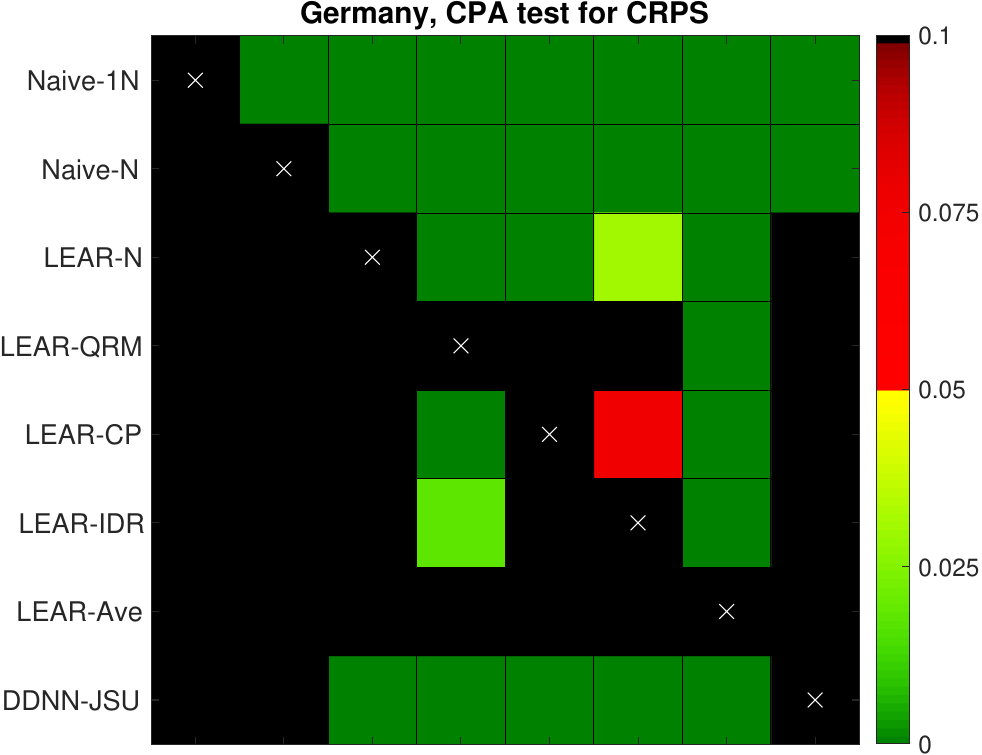}
    \includegraphics[width = 0.4\textwidth]{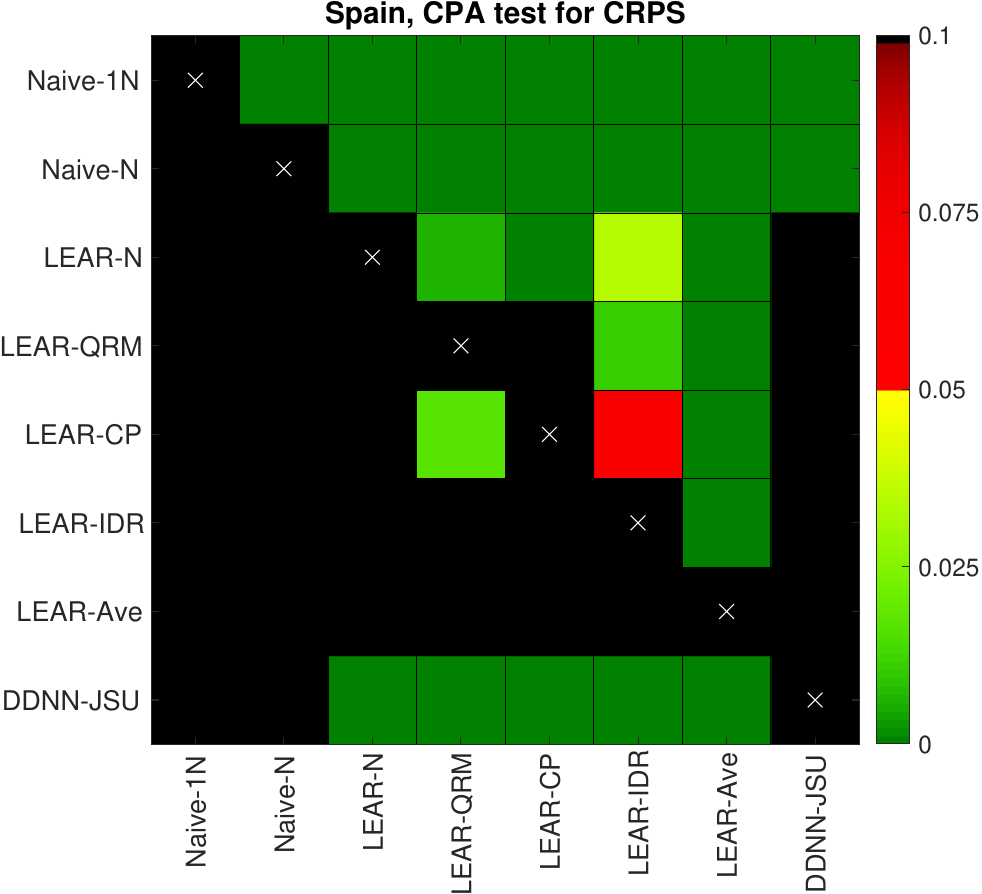}
	\caption{Results of the CPA test of \citet{gia:whi:06} for the CRPS, i.e., the aggregate pinball score for all 99 percentiles, for the whole 4.5-year German (\textit{top}) and Spanish (\textit{bottom}) test period.
    Heat maps are used to illustrate the range of $p$-values -- the smaller they are ($\rightarrow$ dark green), the more significant the difference between the two forecasts (the model on the X-axis outperforms the model on the Y-axis).}
	\label{fig:GW}
\end{figure}

\subsection{Conditional predictive ability}
\label{ssec:Results:GW}

Following \cite{lag:mar:sch:wer:21} and \cite{oli:cha:mar:wer:dub:23}, we run the test of \textit{conditional predictive ability} \cite[CPA;][]{gia:whi:06} to formally assess the performance of the different models. Namely, we test the null $H_0:\boldsymbol{\phi}=0$ in the  regression:   
\begin{equation}\label{eqn:GW}
\Delta_d=\boldsymbol{\phi}' {X}_{d-1} + \epsilon_d, 
\end{equation}
where ${X}_{d-1}$ contains elements from the information set on day $d-1$, i.e., a constant and lags of the loss differential series $\Delta_{d} = ||\varepsilon_{1,d}||_p - ||\varepsilon_{2,d}||_p$, $\varepsilon_{i,d}$ is the $H$-dimensional vector of prediction errors of model $i$ for day $d$,  $||\varepsilon_{i,d}||_p = (\sum_{h=1}^{H} |\varepsilon_{i,d,h}|^p)^{1/p}$ is the $p$-th norm of that vector, and $\epsilon_d$ is an error term. 
Results of the CPA test for all pairs of models (due to the very poor performance reported in Tab.\ \ref{tab:res}, the Naive-1N and Naive-N benchmarks are not considered) and both markets are illustrated in Fig.\ \ref{fig:GW}. Heat maps are used to denote the range of $p$-values -- the smaller they are ($\rightarrow$ dark green), the more significant the difference between the two forecasts (the model on the X-axis outperforms the model on the Y-axis).

Clearly, the LEAR-Ave model significantly outperforms all models; the columns corresponding to this ensemble are dark green in both panels of Fig.\ \ref{fig:GW}. Remarkably, all LEAR-based models, even the LEAR-N benchmark, significantly outperform the DDNN-JSU, a model that is much more complex and computationally much more demanding; see the five dark green cells in the bottom row in both panels. This is a result of the poor performance of the neural network during the energy crisis and the war in Ukraine -- Nov 2021 to Dec 2023 in Germany and Nov 2021 to Mar 2023 in Spain. Potentially, hyperparameter optimization conducted every few months could improve the model's predictive accuracy. This, however, would be a very time consuming task, see Sec.\ \ref{ssec:DDNN:bench}.

\begin{table}[tb]
\caption{Aggregate Pinball Score (APS$_{20}$) for 20 extreme percentiles (1,...,10 and 90,..., 99, i.e., corresponding to confidence levels typically considered in risk management) for the considered models and markets. Like in Table \ref{tab:res}, cells are colored independently for each row and market.
The test period labeled `2000$^\dag$' spans from 27.06.2019 to 31.12.2020 (554 days), the remaining three span full years (365 days).
} 
\label{tab:res:extreme}
\footnotesize
\begin{center}
\begin{tabular}{lrrrr}
\hline
Model    & 2020$^\dag$ & 2021 & 2022 & 2023\\
\hline
         & \multicolumn{4}{c}{\textit{Germany}}\\
Naive-1N & \cellcolor[HTML]{E67C73}1.728 & \cellcolor[HTML]{E67C73}4.804 & \cellcolor[HTML]{E67C73}11.334 & \cellcolor[HTML]{E67C73}5.786 \\
Naive-N  & \cellcolor[HTML]{E88172}1.669 & \cellcolor[HTML]{EB8C70}4.331 & \cellcolor[HTML]{E98471}10.805 & \cellcolor[HTML]{EA8971}5.270 \\
LEAR-N   & \cellcolor[HTML]{FFD366}0.691 & \cellcolor[HTML]{D8CF6F}2.006 & \cellcolor[HTML]{C0CB74}4.629  & \cellcolor[HTML]{FFD666}2.121 \\
LEAR-QRM & \cellcolor[HTML]{A8C879}0.602 & \cellcolor[HTML]{93C47D}1.819 & \cellcolor[HTML]{AFC978}4.579  & \cellcolor[HTML]{9EC67B}1.949 \\
LEAR-CP  & \cellcolor[HTML]{FFD666}0.655 & \cellcolor[HTML]{E6D26C}2.045 & \cellcolor[HTML]{C1CC74}4.631  & \cellcolor[HTML]{F2D369}2.081 \\
LEAR-IDR & \cellcolor[HTML]{F8D568}0.648 & \cellcolor[HTML]{FFD466}2.176 & \cellcolor[HTML]{FFD466}4.985  & \cellcolor[HTML]{87C280}1.914 \\
LEAR-Ave & \cellcolor[HTML]{79C083}0.575 & \cellcolor[HTML]{57BB8A}1.654 & \cellcolor[HTML]{57BB8A}4.327  & \cellcolor[HTML]{57BB8A}1.837 \\
DDNN-JSU & \cellcolor[HTML]{57BB8A}0.555 & \cellcolor[HTML]{F9C169}2.763 & \cellcolor[HTML]{FAC269}6.321  & \cellcolor[HTML]{FDCE67}2.437
\\
\hline
         & \multicolumn{4}{c}{\textit{Spain}} \\
Naive-1N & \cellcolor[HTML]{E67C73}0.981 & \cellcolor[HTML]{E67C73}3.914 & \cellcolor[HTML]{E67C73}5.851 & \cellcolor[HTML]{E67C73}4.184 \\
Naive-N  & \cellcolor[HTML]{E98771}0.920 & \cellcolor[HTML]{EC9170}3.490 & \cellcolor[HTML]{EA8871}5.514 & \cellcolor[HTML]{E98471}4.003 \\
LEAR-N   & \cellcolor[HTML]{FFD566}0.447 & \cellcolor[HTML]{B5CA76}1.870 & \cellcolor[HTML]{F2D369}3.290 & \cellcolor[HTML]{FFD566}2.117 \\
LEAR-QRM & \cellcolor[HTML]{94C47D}0.402 & \cellcolor[HTML]{A7C779}1.841 & \cellcolor[HTML]{AEC978}3.177 & \cellcolor[HTML]{E8D26B}2.045 \\
LEAR-CP  & \cellcolor[HTML]{FFD566}0.446 & \cellcolor[HTML]{CDCE71}1.922 & \cellcolor[HTML]{EAD26B}3.278 & \cellcolor[HTML]{FFD566}2.102 \\
LEAR-IDR & \cellcolor[HTML]{E9D26B}0.431 & \cellcolor[HTML]{FED266}2.130 & \cellcolor[HTML]{FFD666}3.333 & \cellcolor[HTML]{57BB8A}1.856 \\
LEAR-Ave & \cellcolor[HTML]{57BB8A}0.381 & \cellcolor[HTML]{57BB8A}1.671 & \cellcolor[HTML]{57BB8A}3.031 & \cellcolor[HTML]{66BD87}1.876 \\
DDNN-JSU & \cellcolor[HTML]{91C47E}0.401 & \cellcolor[HTML]{FCCB67}2.259 & \cellcolor[HTML]{FCC967}3.696 & \cellcolor[HTML]{89C380}1.922
\\
\hline
\end{tabular}
\end{center}
\end{table}

\begin{figure}[tb]
	\centering 
	\includegraphics[width = 0.4\textwidth]{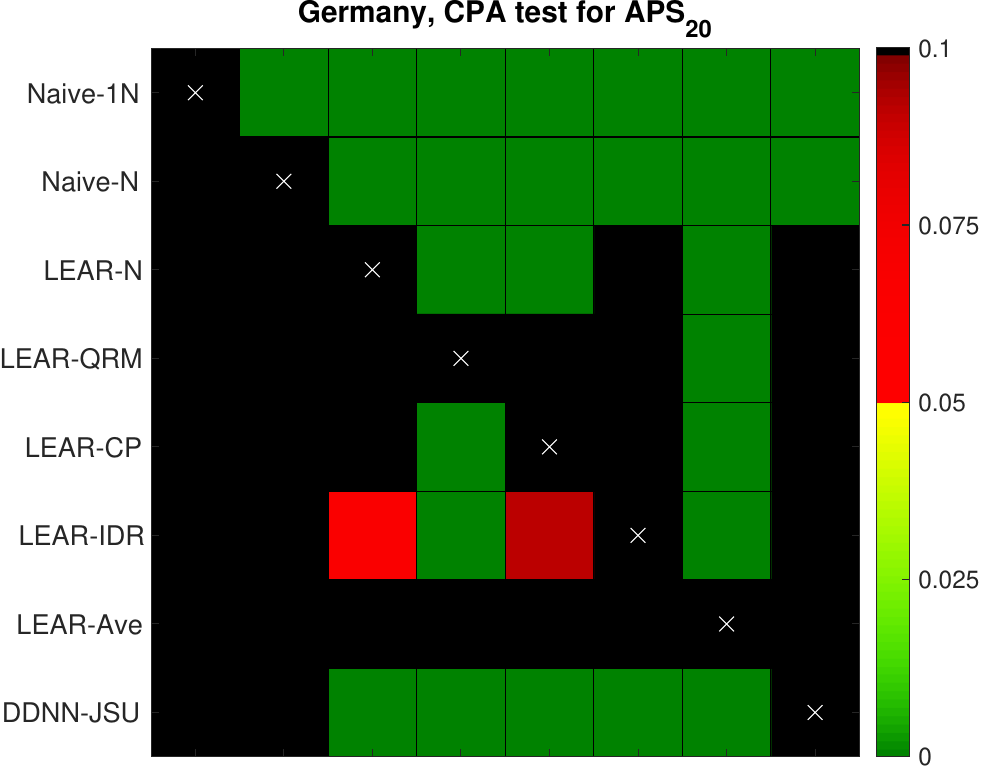} 
    \includegraphics[width = 0.4\textwidth]{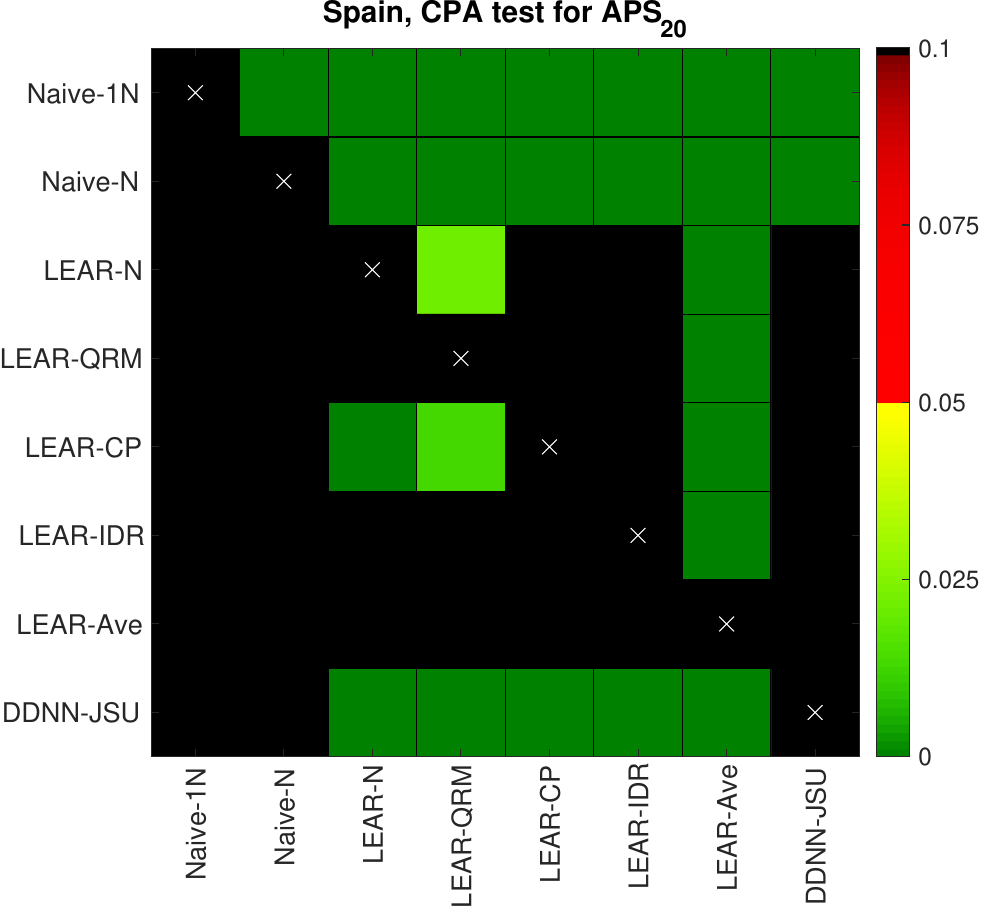}
	\caption{Results of the CPA test of \citet{gia:whi:06} for the APS$_{20}$ metric, i.e., the aggregate pinball score for the extreme top 10 and bottom 10 percentiles, for the whole 4.5-year German (\textit{top}) and Spanish (\textit{bottom}) test periods. The same type of a heat map is used as in Fig.\ \ref{fig:GW}.}
	\label{fig:GW:extreme}
\end{figure}

\subsection{Performance in the tails of the distribution}
\label{ssec:Results:extreme}

In a risk management context we are interested in the tail behavior of the profit and loss (P\&L) distribution. Hence, following \cite{uni:mar:wer:19ENEECO}, we now consider only the percentiles that correspond to confidence levels typically used in risk management: below 10\% and above 90\%, i.e., the lower 10 and the upper 10 percentiles. 
In Table \ref{tab:res:extreme} we report the APS$_{20}$ for the considered models and markets. The corresponding  $p$-values of the CPA test for all pairs of models are illustrated in Fig.\ \ref{fig:GW:extreme}. 

This time, the LEAR-Ave ensemble yields the lowest CRPS across both markets and all test subperiods, except the first subperiod labeled `2000$^\dag$' in Germany (where DDNN-JSU excels, but the difference is statistically insignificant) and year 2023 in Spain (where it is outperformed by LEAR-IDR). Still, the combination significantly outperforms all other models over the entire 4.5-year test period, see the CPA test results in Fig.\ \ref{fig:GW:extreme}. 

Similarly as for the CRPS, all LEAR-based models significantly outperform the DDNN-JSU ensemble across the whole test sets; see the five dark green cells in the bottom row in both panels of Fig.\ \ref{fig:GW:extreme}. Yet, the high accuracy of the neural network in both markets in the first subperiod, i.e., directly after hyperparameter optimization, suggests that the DDNN-JSU has potential, especially in a risk management context. 

Finally, comparing Figs.\ \ref{fig:GW} and \ref{fig:GW:extreme}, we can observe that LEAR-IDR performs better overall (its forecasts are significantly more accurate that those of LEAR-N and LEAR-CP for Germany, and those of all three LEAR-based models for Spain) than for the extreme 20 percentiles (cells in the column corresponding to LEAR-IDR are black in Fig.\ \ref{fig:GW:extreme}). This indicates that IDR is better than its competitors for the more central quantiles.

\section{Conclusions}
\label{sec:Conclusions}

Our study is the first to consider Isotonic Distributional Regression (IDR) and one of the first to use Conformal Prediction (CP) for electricity price forecasting. Overall, it highlights postprocessing as a relatively simple and well-performing means of deriving predictive distributions from point forecasts in such a challenging environment. 

Like \cite{nit:wer:23}, we find that introducing diversity to a pool of forecasts is highly beneficial. Combining the IDR-generated predictive distributions with those of the generally better performing QRA and CP schemes significantly improves the accuracy, as measured by Shapley values. The resulting LEAR-Ave combination outperforms state-of-the-art Distributional Deep Neural Networks of \cite{mar:nar:wer:zie:23} over two 4.5-year test periods from the German and Spanish power markets, spanning the COVID pandemic and the war in Ukraine.

In the tails of the predictive distribution the situation is less straightforward. While for the whole test periods the LEAR-Ave ensemble significantly outperforms the DDNN-JSU model for both markets, as measured by the Conditional Predictive Ability (CPA) test of \cite{gia:whi:06}, in the first 1.5-year subperiod in Germany the DDNN-JSU network excels (the difference is statistically insignificant). Overall, we recommend the LEAR-Ave ensemble as a top performer and the LEAR-QRA model as a powerful all-rounder, second only to the combination. The DDNN-JSU network can provide accurate predictions in the tails of the distribution. However, it is beyond the scope of this study to examine whether frequent (extremely time-consuming) hyperparameter optimization would allow it to perform well during periods of extreme prices.

\section*{Acknowledgments}

The study was partially supported by the National Science Center (NCN, Poland) through grant no.\ 2018/30/A/HS4/00444 (to AL), grant no.\ 2023/49/N/HS4/02741 (to BU) and grant no.\ 2021/43/I/HS4/02578 (to RW).

\bibliographystyle{elsarticle-harv}
\bibliography{ref}

\end{document}